\documentclass[]{qjrms4}

\usepackage[dvips,colorlinks,bookmarksopen,bookmarksnumbered,citecolor=red,urlcolor=red]{hyperref}

\newcommand\BibTeX{{\rmfamily B\kern-.05em \textsc{i\kern-.025em b}\kern-.08em
T\kern-.1667em\lower.7ex\hbox{E}\kern-.125emX}}

\usepackage{moreverb}

\def\volumeyear{2010}

\usepackage{natbib}
\newcommand{\pdf}{\textsc{PDF}}
\renewcommand{\vec}{\boldsymbol}
 \newcommand{\qs}{q_{\mathrm{s}}}
  
  \newcommand{\ys}{y_{\mathrm{s}}}
\newcommand{\dd}{\mathrm{d}}
\newcommand{\ee}{\mathrm{e}}
\renewcommand{\S}{\mathcal{S}}
\newcommand{\C}{\mathcal{C}}
\newcommand{\F}{\mathcal{F}}
\newcommand{\qmin}{q_{\mathrm{min}}}

\newcommand{\qmax}{q_{\mathrm{max}}}
\newcommand{\rmin}{r_{\mathrm{min}}}
\newcommand{\tmoist}{\tau_{\mathrm{Moist}}}

\newcommand{\la}{\langle}
\newcommand{\ra}{\rangle}

\newcommand{\defn}{\stackrel{\mathrm{def}}{=}}

\begin{document}

\runningheads{J. Sukhatme and W. R. Young}{Advection-Condensation}

\title{The advection-condensation model and water vapour \pdf{s}}

\author{\corrauth Jai Sukhatme\affil{a}   William R. Young\affil{b} }

\address{\affilnum{a} Centre for Atmospheric and Oceanic Sciences, Indian Institute of Science, Bangalore 560012, India \\
\affilnum{b} Scripps Institute of Oceanography, University of California at San Diego, San Diego, CA 92093-0213, USA.}

\corraddr{jai.goog@gmail.com}

\begin{abstract}

The statistically steady humidity distribution resulting from an interaction of advection, modeled as an uncorrelated random walk of moist parcels on an isentropic surface, and a vapour sink, modeled as immediate condensation whenever  the specific humidity  exceeds a specified saturation humidity, is explored with theory and simulation.   
A source supplies moisture at the deep-tropical southern boundary of the domain, and the saturation humidity is specified as a monotonically decreasing function 
of distance from the  boundary. The boundary source  balances the interior condensation sink,  so that a stationary 
spatially inhomogeneous humidity distribution emerges. An exact solution of the Fokker-Planck equation delivers a simple expression for
the resulting  probability density function (\pdf{}) of the water vapour field and also of the relative humidity. 
This solution agrees completely with a numerical simulation of the process, and the humidity \pdf{} exhibits several features of interest, 
such as bimodality close to the source and unimodality further from the source. The \pdf{}s of specific and relative humidity are broad and non-Gaussian.
The domain averaged relative humidity \pdf{} is bimodal with distinct moist and dry peaks, a feature which we show  agrees with
middleworld isentropic \pdf{}s derived from the ERA
interim dataset.

\end{abstract}


\maketitle


\section{Introduction} 

Water vapour plays a significant role in diverse problems pertaining to the dynamics of
the Earth's climate \citep{RayNature, SGL-2010}. In particular, with regards to its radiative effects, estimating the
amount of water vapour in the upper troposphere is crucial \citep{HS, SB}. 
Significant efforts --- especially directed
towards the subtropical troposphere --- have yielded a key insight into the nonlocal nature
of processes that control the distribution of water vapour 
\citep{YP-1994, Sher, SH-1997, P-1998, RP-1998, GSH-2005, BRP-2009}.

The resultant framework
for understanding the evolution of water vapour, or generally any condensable
substance in a fluid dynamical setting, is known as the advection-condensation
(AC) model.
The recent review by \cite{Sher-rev} provides an introduction, while the book chapter by \cite{PBR}  (hereafter, PBR) 
provides a thorough overview 
of the AC model.

Further, as elaborated on by \cite{SB} and PBR, in addition to the amount, the probability density function (\pdf{}) of the water vapour field has a strong influence
on the longwave cooling of the atmosphere.
Explicit calculations illustrating the effect of the \pdf{} on the outgoing longwave radiation (OLR) are given by \cite{ZSM}. The importance of the \pdf{} 
stems from the approximately logarithmic dependence of the change in OLR to the specific humidity of the water vapour in the domain. Further discussion of
this  logarithmic 
dependence which results from a spectral broadening of absorption peaks can be found in the text by  \cite{Ray-book}. 

Following \cite{SB-1993}, there have been numerous efforts that examine water vapour \pdf{}s in the tropical and subtropical
troposphere. A principal feature noted by these studies is the non-normality of the \pdf{}s. A range of distributions, 
from bimodality in the deep tropics \citep{BZ-1997, ZSM, MF-2006, LKJ-2007}, to unimodal \pdf{}s with a roughly lognormal or power-law form in the 
subtropics  have been documented \citep{SB-1993, SKR}. \cite{Ryoo} have emphasized that a salient and baffling 
feature of water vapour \pdf{}s is their spatial inhomogeneity. 
Models based on the statistics of subsidence drying and random  re-moistening of parcels  reproduce many of the observed features of the  humidity \pdf{}s, 
provided that one considers the model parameters to be functions of position determined to match observations \citep{SKR, Ryoo}. 

One of the main achievements of this paper is an exact solution for the water vapour PDF that characterizes the statistically 
steady state emerging from an interaction of advection, 
condensation and a spatially localized moisture source. 
In particular, we consider  moist parcels advected by a white-noise stochastic velocity field. 
Each parcel carries a specific humidity and experiences condensation whenever and wherever the specific humidity exceeds a specified saturation profile. 
The parcel humidity is reset to saturation by a moist source at the southern boundary of the domain. 
In the ultimate statistically steady state, the southern vapour source is balanced by condensation throughout the interior of the domain. 
In contrast to earlier considerations of the AC model with  white-noise advection  \citep{GS-2006}, we consider statistically steady states, 
rather than the initial value problem, and we obtain the full water-vapour \pdf{} rather than the mean humidity.

The AC model is described in more detail  in sections \ref{model} and \ref{brownian}, and in section \ref{solution}
we  solve the Fokker-Planck equation to obtain the \pdf{} of the specific humidity. The utility of the \pdf{} is demonstrated by calculating the 
average moisture flux and condensation. In section \ref{example} we focus on a particular saturation profile --- the standard exponential model ---  and exhibit  analytical solutions for the specific and relative 
humidity \pdf{}s; these are  shown to be in very good agreement with \pdf{}s estimated from a Monte Carlo simulation. 
In section \ref{discussion}  we analyse daily
isentropic relative humidity data from the ERA interim product, and show that the \pdf{}s derived from this dataset  agree with principal features
predicted by our idealized model.
A discussion of the \pdf{}s, with implications for the 
atmospheric radiation budget and the present-day distribution of water vapour, followed by a conclusion section \ref{conclusion}  ends the paper.

\section{The AC model \label{model}}

The central quantity in the AC model is the specific humidity, defined by
\begin{equation}
q(\vec{x},t) \defn \frac{\textrm{mass of water vapour in a parcel}}{\textrm{total mass of the parcel}}\, ,
\label{Da}
\end{equation}
where $\vec{x}=(x,y)$ is position on an isentropic surface.
Our main focus here is the extratropical vapour distribution, which is determined in large part by the wandering of air parcels on isentropic 
surfaces and the temperature changes associated with latitudinal excursions.

Following PBR, the AC model is:
\begin{equation}
\partial_t q + \vec{u \!\cdot \! \nabla } q = \S - \C\, .
\label{D1}
\end{equation}
Above, $\S$ is the vapour source,  $\C$ is the condensation sink and $\vec{u}(\vec{x},t)$ is  velocity on an isentropic surface. In the absence of sources and sinks, $Dq/Dt=0$, i.e.\ the specific humidity is a materially conserved quantity. The AC formalism does not
consider the diffusive homogenization of water vapour e.g., via the addition of  $\chi \nabla^2 q$ on the right of \eqref{D1}. This  limitation of the  model is discussed further in section \ref{discussion}. 

Latent heat 
release and radiative cooling result in significant cross-isentropic motion in the troposphere, and we have ignored this process in formulating a two-dimensional  isentropic model in \eqref{D1}.  We emphasize that the  AC formulation is not
inherently limited to two-dimensions or to motion on isentropic surfaces. But for ease of exposition, and simplicity, isentropic motion is a useful starting point.

The condensation sink on the right of \eqref{D1} might be modeled with 
\begin{equation}
\C = \tau^{-1}_c\, \left[q(\vec{x},t)-\qs(\vec{x})\right]  \, H\left[q(\vec{x},t)-\qs(\vec{x})\right]\, ,
\label{D2}
\end{equation}
where $\tau_c$ is the timescale associated with condensation, $\qs(\vec{x})$ is the prescribed saturation specific humidity; 
$H(x)$ denotes the Heaviside step function, which is zero if the $x<0$ and one if $x>0$. 
Observed atmospheric supersaturations rarely exceed a few percent and therefore the sink $\C$ is extremely effective \citep{WH}. 
In other words, we are concerned with the rapid-condensation limit $\tau_c \to 0$. 
In the rapid-condensation limit, the details in $\C$ are unimportant and condensation can instead be represented as a rule enforcing subsaturation:
\begin{equation}
q(\vec{x},t) \to \min\left[ q(\vec{x},t), \qs(\vec{x})\right]\, .
\label{rule}
\end{equation}
Therefore, $q(\vec{x},t) \le \qs(\vec{x})$ and there is an upper  bound on the specific humidity at every $\vec{x}$. 
Globally, the maximum tolerable  specific humidity  is
 \begin{equation}
 \qmax \defn  \max_{\forall \vec{x}} \qs(\vec{x})\, .
 \label{tolerable}
 \end{equation}

If the flow is ergodic then a wandering parcel explores the entire domain and 
eventually encounters the dryest regions, with the smallest  value of $\qs(\vec{x})$. Thus
in the absence of sources \citep{SP-unpub}
\begin{equation}
\lim_{t \to \infty} q(\vec{x},t) = \qmin\, ,
\end{equation} 
where
\begin{equation}
\qmin \defn \min_{\forall \vec{x}} \qs(\vec{x})\, .
\end{equation}
On the other hand, without advection, the source will eventually saturate $q(\vec{x},t)$ at every point $\vec{x}$ where $\S \neq 0$. Therefore, interesting statistically steady states are  expected only if there is both a source $\S$  and  advection  $\vec{u}(\vec{x},t)$. 

The problem is to determine the evolution and the statistics of the specific humidity $q(\vec{x},t)$, given the velocity field $\vec{u}(\vec{x},t)$ and the  saturation profile $\qs(\vec{x})$. 
Due to its physical significance, the principal statistical feature we focus here on is the probability density function (\pdf{}) of $q$.
Quantities such as the mean moisture flux and the mean condensation rate are also crucial, and have been examined in the AC framework \citep{GS-2006,GLS-2011}. 
These mean  quantities can  be deduced from \pdf{} of $q$ via integration over $q$.

\section{Moist brownian parcels \label{brownian}}

\begin{figure}
\centering
\includegraphics[width=7cm,height=6cm]{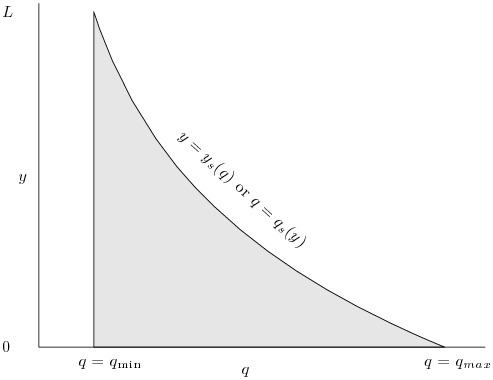}
\caption{In the rapid-condensation limit the accessible domain in the $(q,y)$ plane is the shaded region shown above. The model assumes that $\qs(y)$ decreases monotonically from $\qmax = \qs(0)$ to   $\qmin = \qs(L)$. The functions $\qs(y)$ and $\ys(q)$ are inverses of each other: $\qs(\ys(q))=q$.}
\label{figa}
\end{figure}

The simplest representation of random transport is obtained by taking the velocity $\vec{u}(\vec{x},t)$ in \eqref{D1} as having rapid temporal decorrelation so that the motion of an air parcel is brownian. Then the spatial evolution of an ensemble  of parcels, each carrying its own individual value of $q$,  is described by the diffusion equation. We also assume that $q_s$ depends only on the meridional coordinate $y$ in the domain $0<y<L$, and that $\qs$ decreases monotonically from $\qmax$ at $y=0$ to $\qmin$ at $y=L$. Each parcel is located by a point in the configuration space $(q,y)$.  Because of the condensation rule \eqref{rule}, the ensemble of parcels is confined to the shaded region of configuration space illustrated in Figure \ref{figa}. 

The AC model is then expressed as a set of coupled stochastic differential equations
governing the position $Y(t)$ and specific humidity $Q(t)$ of a moist brownian parcel
\begin{align}
\dd Y(t) &= \sqrt{2 \kappa_b} ~\dd W(t), \nonumber \\
\dd Q(t) &= \left\lbrace  \S\left[Y(t)\right] - \C\left[Y(t),Q(t)\right] \right\rbrace ~\dd t
 \label{1aa}
\end{align}
where $W(t)$ is a Weiner process,  $\kappa_b$ is the brownian diffusivity associated with the random walk, and $\S$ and $\C$ are the source and condensation functions in \eqref{D1}. The numerical simulations described below amount to Euler-Maruyama discretization of \eqref{1aa}. The 
midlatitude baroclinic eddies responsible for transport have non-trivial spatio-temporal correlations \citep{Sukh-2005, GS-2006}. Thus the white noise advection used in \eqref{1aa} and in earlier AC  studies is an idealization.

The statistical properties of the ensemble of moist brownian parcels are characterized by a PDF $P(q,y,t)$, such that the expected number of parcels in $\dd q \,  \dd y$ is equal to
\begin{equation}
N\, P(q,y,t) \, \dd q \,  \dd y \, ,
\end{equation}
where $N$ is total number of parcels.

The evolution of $P(q,y,t)$ is  governed by the  Fokker-Planck equation
\begin{equation}
\partial_t P + \partial_q [(\S - \C)P] = \kappa_b \partial^2_{y} P\, ,
\label{1a}
\end{equation}
with the normalization,
\begin{equation}
\int_{0}^{L} \int_{\qmin}^{\qmax} \!\!\! P(q,y,t) \, \dd q \, \dd y  = 1\, . 
\label{2a}
\end{equation}

In the limit of rapid condensation, as $\qmin \le q(y,t) \le q_s(y)$, the normalization takes the form
\begin{equation}
\int_{0}^{L} \int_{\qmin}^{\qs(y)} \!\!\! P(q,y,t) \, \dd q \, \dd y  = 1\, . 
\label{2aa}
\end{equation}
Integrating $P(q,y,t)$ over $q$ produces the marginal density
\begin{equation}
p(y,t) \defn \int_{\qmin}^{\qs(y)} \!\!\! P(q,y,t) \, \dd q\, .
\label{marg}
\end{equation}
Because the brownian motion of a parcel is independent of the specific humidity it carries, we expect that integration of \eqref{1a} over $q$ will show that 
$p(y,t)$ satisfies the diffusion equation, and indeed this result is immediate
\begin{equation}
\partial_t p = \kappa_b \partial_{y}^2 p\, .
\label{diffeq}
\end{equation}
If parcels are reflected back into the domain at $y=0$ and $L$ then    the solution of \eqref{diffeq} is  $\lim_{t \to \infty} p(y,t) = L^{-1}$. Thus if we seek  long-time equilibrium solutions of the Fokker-Planck equation \eqref{1a}, then we require 
\begin{equation}
L^{-1} = \int_{\qmin}^{\qs(y)} \!\!\! P(q,y) \, \dd q\, .
\label{prominent}
\end{equation}
Note that the left of \eqref{prominent} is consistent with the global normalization in \eqref{2a}.
In the limit $\tau_c \to 0$,  the steady, marginal normalization condition \eqref{prominent} applies at every $y$, 
and for all models of $\S$ and  $\C$. 
In other words, in  a statistical steady state the  normalization condition \eqref{prominent} is equivalent to enforcing the rapid-condensation rule in \eqref{tolerable}.  The normalization \eqref{prominent} figures prominently in the following solution of the Fokker-Planck equation. Similar ``variable limit" normalization conditions are encountered  in the solution of
integrate-and-fire neuron models \citep{FM-1998}.

\section{Steady solution  with  ``resetting'' at $y=0$ \label{solution}}

Now we adopt a ``resetting'' model for the vapour source $\S$:
 on encountering the southern boundary at $y=0$, parcels are reflected back into the domain with a new value of $q$ chosen from a specified probability density function $\Phi(q)$, with the normalization
\begin{equation}
\int_{\qmin}^{\qmax} \!\!\!\!\!\! \Phi(q) \, \dd q = 1\, .
\label{B1}
\end{equation}
For example, completely saturating the parcels at $y=0$ corresponds to the choice $\Phi(q)=\delta(q-\qmax)$. This is a simple representation of tropical moistening. Resetting boundary 
conditions have been employed in a previous study of AC on isentropic surfaces \citep{YP-1994}, as well as in advection-diffusion of
passive scalars \citep{Ray-csf, NHP}. 

We emphasize  that resetting the humidity only at $y=0$ is a strong model assumption that  distinguishes this work from the AC models developed by \cite{SKR} and \cite{Ryoo}. 
The statistical model developed in those works assumes that random resetting occurs throughout the domain as an exponential or gamma stochastic process with a characteristic 
time scale $\tmoist(\vec{x})$. In effect, by resetting only at $y=0$, we are considering the extreme case in which $\tmoist(\vec{x}) = \infty$, except at $y=0$. 
Physically, the picture we have in mind is that parcels experience moistening when they encounter 
convective regions which are restricted to the tropics, and therefore
$y$ is the distance from this saturated zone. In fact, this serves as another limiting case for the AC model along with the re-setting 
throughout the domain scenario considered by \cite{SKR} and \cite{Ryoo}. 

An important consequence of resetting only at $y=0$ is that the dryest parcels,  
which are created at $y=L$ with $q=\qmin$, maintain their extreme dryness till they eventually strike $y=0$. 
The solution below requires explicit consideration of these dryest parcels by inclusion of a component $\delta(q-\qmin)$, referred to as the ``dry-spike'', in the \pdf. 

\subsection{Solution of the Fokker--Planck equation}
In the rapid-condensation limit, the accessible part of the $(q,y)$-space is the shaded area in Figure \ref{figa}, where $\C=0$. But with the resetting boundary condition, the source $\S$ is also zero within the shaded region. Thus the steady Fokker-Planck equation in the shaded region of Figure \ref{figa} collapses to $\partial_{y}^2 P=0$, with the immediate solution
\begin{equation}
P(q,y) = \frac{\Phi(q)}{L} + y B(q)  \, .
\label{4}
\end{equation}
The first term on the right of \eqref{4} is determined by the resetting condition and by application of the normalization in \eqref{prominent} at $y=0$. 

Corresponding to the dryest parcels in the domain, $B(q)$ in \eqref{4} must contain a component proportional to $\delta(q-\qmin)$. This ``dry spike''  contains parcels  created by encounters with the northern boundary at $y=L$. These considerations refine \eqref{4} to
\begin{equation}
P(q,y) = \frac{\Phi(q)}{L} + \left[ \frac{\delta(q-\qmin)}{L^2} + F(q)\right] y\, .
\label{5}
\end{equation}
The term $ L^{-2} \delta(q-\qmin)$ in \eqref{5} is obtained by applying the normalization condition \eqref{prominent} at $y=L$. 

In \eqref{5},  $F(q)$ is the smooth part of the PDF, which is determined by substituting into the normalization  \eqref{prominent}. This requirement leads to 
\begin{equation}
\frac{L}{y} \int_{q_s(y)}^{\qmax} \!\!\!\!\!\! \Phi(q') \, \dd q' = 1 + L^2 \int_{\qmin}^{q_s(y)} \!\!\! F(q') \, \dd q'\, . 
\label{B4}
\end{equation}
Inspecting (\ref{B4}), we see that it is possible to solve for $F(q)$ by using $q$, rather than $y$, as the independent
variable. That is, following the notation in Figure \ref{figa}, we use $y=y_s(q)$ to re-write (\ref{B4}) as 
\begin{equation}
\frac{L}{y_s(q)} \int_{q}^{\qmax} \!\!\!\!\!  \Phi(q') \, \dd q' = 1 + L^2 \int_{\qmin}^{q} \!\!\! F(q') \, \dd q' \, .
\label{B5}
\end{equation}
The derivative with respect to $q$ then determines $F(q)$, i.e.\ 
\begin{equation}
L F(q) = \frac{d}{dq} \frac{\Lambda(q)}{y_s(q)}  \, ,
\label{B6}
\end{equation}
where $\Lambda(q)$ is the cumulative distribution
\begin{equation}
\Lambda(q) \defn \int_{q}^{\qmax} \!\!\!\!\!\! \Phi(q') \, \dd q' \, .
\label{Lambdadef}
\end{equation}

Assembling  the pieces,  we obtain the \pdf
\begin{equation}
P(q,y) = 
 - \frac{1}{L} \frac{\dd \Lambda}{\dd q} + \frac{y}{L}  \left[ \frac{\delta(q-\qmin)}{L} +  \frac{d}{dq} \frac{\Lambda(q)}{y_s(q)}\right] \, .
\label{6}
\end{equation}
The expression above applies only within the shaded region in Figure \ref{figa} where $q<\qs(y)$; in the supersaturated region $q>\qs(y)$, $P(q,y)=0$. The \pdf{} is discontinuous at the saturation boundary $q=\qs(y)$. This completes the solution
for the steady-state Fokker-Planck equation with resetting forcing at $y=0$. 

In summary, in the rapid-condensation limit, and with brownian motion,  the   steady   Fokker-Planck equation is solved exactly by \eqref{6}.  The normalization constraint in  \eqref{prominent}, and the use $q$ as the independent variable, produces  this relatively simple solution. In fact, the normalization constraint  \eqref{prominent} can also be regarded as a differential boundary condition. Manipulating this boundary condition provides an
alternate route to the solution  outlined in Appendix A.

\subsection{The global \pdf}

The global \pdf{}  of specific humidity is the marginal density
\begin{equation}
g(q) \defn \int_{0}^{\ys(q)} \!\!\!\! P(q,y) \, \dd y \, .   
 \label{global1} 
\end{equation}
The integral on the right can be evaluated using the solution in \eqref{6}, and one finds
\begin{equation}
g(q) = \frac{1}{2} \delta (q - \qmin) - \frac{1}{2} \frac{\dd }{\dd q} \left( \ys(q) \Lambda(q) \right)\, .
\label{global2}
\end{equation}
A comforting check on \eqref{global2} is that
$g(q)$ satisfies the normalization 
\begin{equation}
\int_{\qmin}^{\qmax} \!\!\!  g(q) \, \dd q = 1\, ,
\end{equation}
which is consistent with \eqref{2aa}. 

Independent of model details, such as the specification of $\qs(y)$ and $\Lambda(q)$, the dry spike,  $\delta(q-\qmin)$ in \eqref{global2}, contains 
exactly half of the parcels in the ensemble. Prof.\ P.\ O'Gorman noted that this feature can be understood on the basis of symmetry: exactly half the parcels  have 
more recently encountered the 
moist southern boundary than the dry northern boundary. The former half have  $q \ge \qmin$, while the other half constitute the dry spike with  $q=\qmin$.

\subsection{Averages}

Knowledge of the stationary \pdf{} allows us to evaluate the average of any function of $q$, say $f(q)$, as
\begin{equation}
\langle f \rangle(y)  \defn L \int_{\qmin}^{\qs(y)} \!\!\!\!\! f(q) P(q,y) \, \dd q\, .
\end{equation}
Using the solution for $P(q,y)$ in \eqref{6}, one can show that the average defined above is 
\begin{equation}
\langle f \rangle   =f(\qs) \Lambda(\qs)   - \int_{\qmin}^{\qs} \!\!\!\!\!  f(q) \frac{\dd \Lambda}{\dd q} \, \dd q 
 - y \int_{\qmin}^{\qs} \! \frac{\dd f}{\dd q} \, \frac{\Lambda(q) }{\ys(q)}  \, \dd q\, .
\label{avg1}
\end{equation}
Taking the derivative of \eqref{avg1} with respect to $y$ we obtain, after some  remarkable cancellations,
\begin{equation}
\frac{\dd \langle f \rangle}{\dd y} = - \int_{\qmin}^{\qs} \frac{\dd f}{\dd q} \frac{\Lambda(q)}{\ys(q)} \, \dd q\, .
 \label{avg2}
\end{equation}

\subsection{Average humidity and  condensation rate}

To understand the transport of moisture in this model, and the associated mean condensation rate, we return to \eqref{1a} and momentarily retreat  from the rapid-condensation limit\footnote{We continue using  resetting forcing, so that  $\S=0$ when $y>0$.}. Then, multiplying  \eqref{1a} by $q$ and integrating from $\qmin$ to $\qmax$, we have 
\begin{equation}
- \int_{\qs}^{\qmax} \!\!\!\!\! q \, \partial_q (\C P)\,  \dd q = \kappa_b \, \frac{\dd^2 \la q \ra}{\dd y^2}\, ,
\label{condy1}
\end{equation}
where the condensation sink $\C$ is given by the model in \eqref{D2}. Using this expression for $\C$ and
integrating the left of \eqref{condy1} by parts, we obtain the mean condensation rate, defined by 
\begin{equation}
\la \C \ra  \defn \tau_c^{-1}\int_{\qs}^{\qmax} \!\!\!\!\!\! \left(q - \qs\right) P(q,y)\, \dd q \, ,
\end{equation}
as
\begin{equation}
\la \C \ra = \kappa_b \frac{\dd^2 \la q \ra}{\dd y^2}\, .
\label{condy11}
\end{equation}
The right hand side of \eqref{condy11}  is the convergence of the moisture flux, which balances the mean condensation $\la \C\ra$ on the left.  

This argument shows that the flux of moisture has the same  diffusivity $\kappa_b$ as that of a brownian parcel. 
As explained by  \cite{GS-2006}, this is a special property of the brownian limit: if the velocity has a nonzero integral time scale (e.g., as in the Ornstein-Ulenbeck process) then the 
diffusivity of moisture will be systematically less than the random-walk diffusivity. 
We emphasize that while the moisture flux is diffusive, even with a Brownian advecting flow,
the condensation sink
cannot be adequately represented by a bulk diffusivity. This inadequacy of a mean-field representation of condensation is clearly demonstrated 
in PBR. 

In the rapid condensation limit, the right of \eqref{condy11} can be computed by putting $f(q)=q$ in \eqref{avg2}. Thus if $\tau_c \to 0$,  the mean condensation rate is 
\begin{equation}
\la \C \ra = - \frac{\kappa_b}{y} \frac{\dd \qs}{\dd y}\,  \Lambda\left(\qs(y)\right)\, .
\label{condy2}
\end{equation}
Although $\C \propto \tau_c^{-1} \to \infty$, the mean condensation $\la \C\ra$ in \eqref{condy2} is independent of $\tau_c$. This singular limit is achieved  because the nonzero probability of supersaturation is confined to a boundary layer, lying just above the saturation boundary $q=\qs(y)$ in Figure \ref{figa}. 

If the resetting \pdf{} $\Phi(q) $ is non-singular at $\qmax$,  then $\la \C \ra$ in \eqref{condy2} is finite  as $y\to 0$, despite the factor $y^{-1}$. Specifically, one can show that $\lim_{y \to 0} \Lambda(\qs)/y=- \Phi(\qmax) (\dd\qs/\dd y)$. 

\section{An example \label{example}}

The solution from the previous section is most simply  illustrated by supposing that the resetting produces
complete saturation at $y=0$,
\begin{equation}
\Phi(q)=\delta(q-\qmax)\, ,
\label{delta}
\end{equation}
so that $\Lambda(q)=1$. For the saturation humidity we use the exponential model 
\begin{equation}
\qs(y) = \qmax \,  \ee^{-\alpha y}\, .
\label{exp}
\end{equation}
This form of $\qs$ has been
employed previously in AC studies \citep{PBR, GS-2006}. 

With $\Phi(q)$ and $\qs(y)$ in \eqref{delta} and \eqref{exp}, the steady solution of the Fokker-Planck equation is
\begin{equation}
P(q,0) = \frac{\delta(q-\qmax)}{L} \, ,
\label{B8.1}
\end{equation}
and away from the southern boundary
\begin{equation}
P(q, 0<y ) = \frac{y}{L} \left[\frac{\delta(q-q_{\textrm{min}})}{L} + \frac{\alpha}{q \log^2(q/\qmax)} \right]\, .
\label{B8.2}
\end{equation}
The solution \eqref{B8.2} can be non-dimensionalized with 
\begin{equation}
\alpha_* \defn \alpha L\, , \qquad q_*\defn q/\qmax\, , \qquad y_* \defn  y/L\, ,
\end{equation}
and $P_*(q_*,y_*) = \qmax L  P(q,y)$.
This scaling identifies the fundamental non-dimensional control parameter $ \alpha_* $, 
and the scaled \pdf{} is
\begin{equation}
 P_*(q_*,0<y_*) = y_* \left[ \delta(q_*- \epsilon)  + \frac{\alpha_* }{q_* \log^2(q_*)} \right]\, ,
 \label{starpdf}
\end{equation}
with 
\begin{equation}
\epsilon \defn \frac{\qmin}{\qmax} = \ee^{-\alpha_*}\, .
\label{epsilondef}
\end{equation}
At the southern boundary $P_*(q_*,0)= \delta(q_*-1)$. The main point is that the shape of $P_*(q_*,y_*)$ is controlled by a single parameter $\alpha_*$. 

\begin{figure}
\centering
\includegraphics[width=8cm,height=7cm]{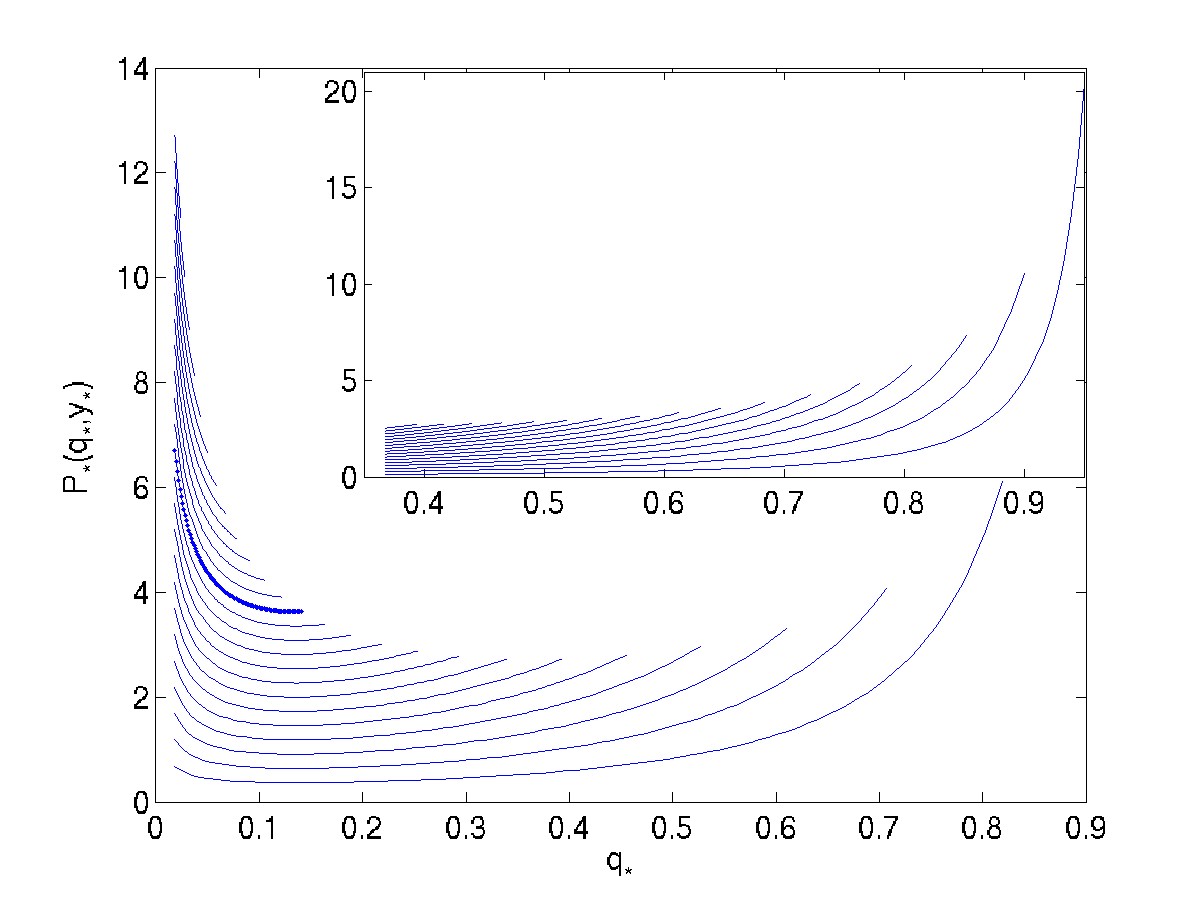}
\caption{\label{fign} The behaviour of the smooth portion of $P_*(q_*,y_*)$ for two choices of $\alpha_*$; we do not show the dry spike at $\qmin$. The major axes correspond to $\alpha_*=4$; there is  a transition from a bimodal to unimodal \pdf{} at $y_*=2/\alpha_*=0.5$. Larger $y_*$'s have smaller ranges of permissible $q_*$, and hence the curves are ordered such that $y_*$ increases  as we move to the left. The  transition curve at $y_*=2/\alpha_*=0.5$ is shown by the dark line in the plot. 
The inset corresponds to $\alpha_*=1$. The 
curves are ordered as in the main plot. In this case, counting the dry spike at $\qmin$,  the \pdf{} is bimodal at all values of $y_*$.}
\end{figure}

There is an interesting change in the structure of $P_*$ at $\alpha_*=2$: 
 $P_*$ is bimodal for all $y_*$ if $\alpha_*<2$; if $\alpha_*>2$ then  $P_*$ is bimodal only in the region $0<y_*<2/\alpha_*$.  This  is illustrated in Figure \ref{fign} which shows the smooth portion of $P_*$ with $\alpha_*=4$ and $\alpha_*=1$. The \pdf{} with $\alpha_*=4$ changes from a bimodal 
to a unimodal form as one crosses $y_*=2/\alpha_*=0.5$. Indeed, for $y_* > 0.5$ the $\alpha_*=4$ \pdf{} falls  monotonically from $q_*=\epsilon$. 
In the case with  $\alpha_*=1$, the \pdf{} increases  with increasing $q_*$ at all values of $y_*$. Therefore, in 
combination with $\delta(q_*- \epsilon)$, the \pdf{} is always bimodal if $\alpha_*<2$.

On a middleworld isentrope, 
the saturation mixing ratio usually changes by more than an order of magnitude as one spans the 
entire midlatitudes, hence from (\ref{exp}), we expect that $\alpha_* > 2$ is the relevant atmospheric case. 
Therefore, close to the vapour source the \pdf{} is bimodal. One mode, the dry spike $\delta(q-\qmin)$, corresponds to a singularity at $\qmin$, after which the \pdf{} falls off, only to rise again to a 
secondary moist  peak as $q \rightarrow \qs(y)$. 
Further from the vapour source (i.e.\ for $y_* > 2/\alpha_*$) the \pdf{} is unimodal with a dry spike at $\qmin$ followed by a gradual 
monotonic decay to a minimum at $q=\qs(y)$.
 
We proceed to discuss the implications  of the solution in \eqref{B8.2} using dimensional variables. The non-dimensional versions of all formulas  
are obtained with  $L\to 1$, $\alpha \to \alpha_*$ and $\qmin \to \epsilon$.

\subsection{Statistics in a strip $y_1<y<y_2$}

\begin{figure}
\centering
\includegraphics[width=8cm,height=7cm]{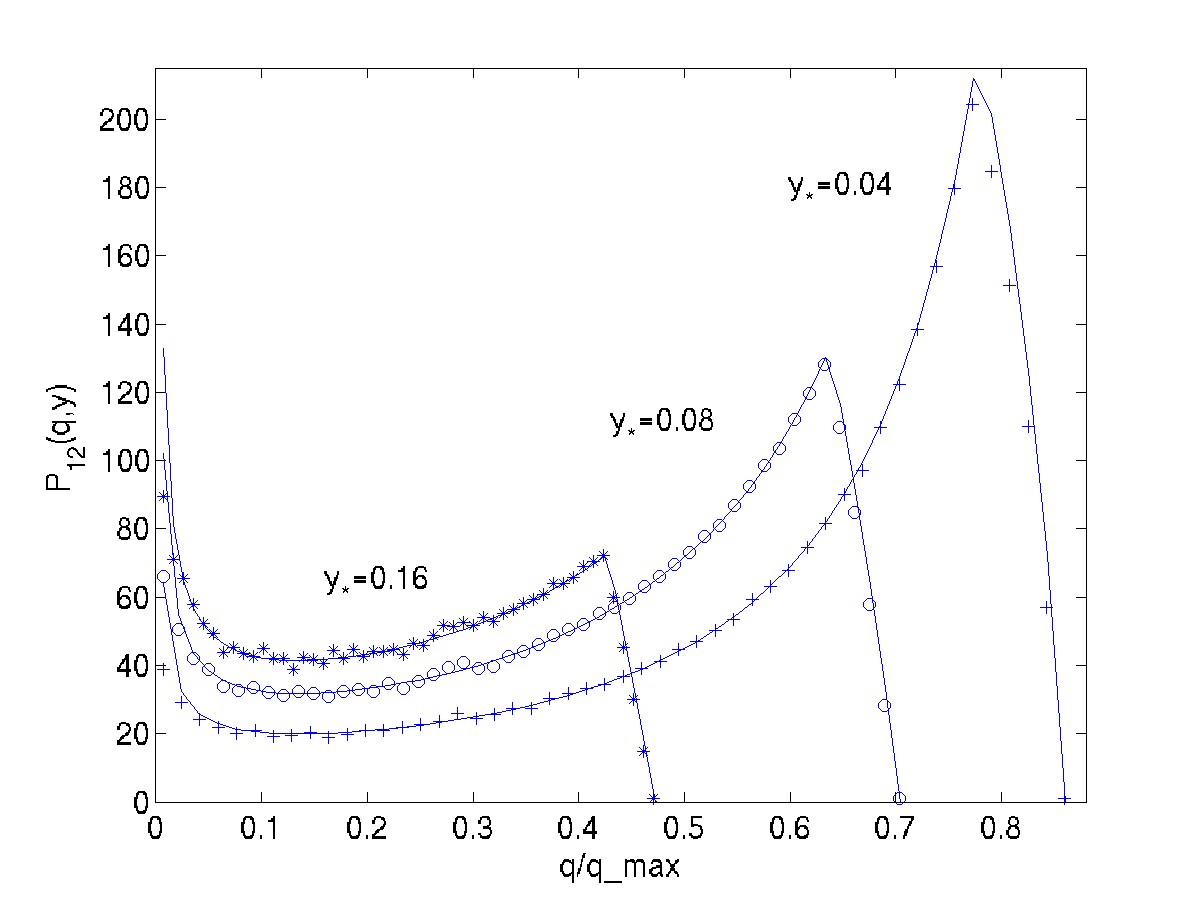}
\includegraphics[width=8cm,height=7cm]{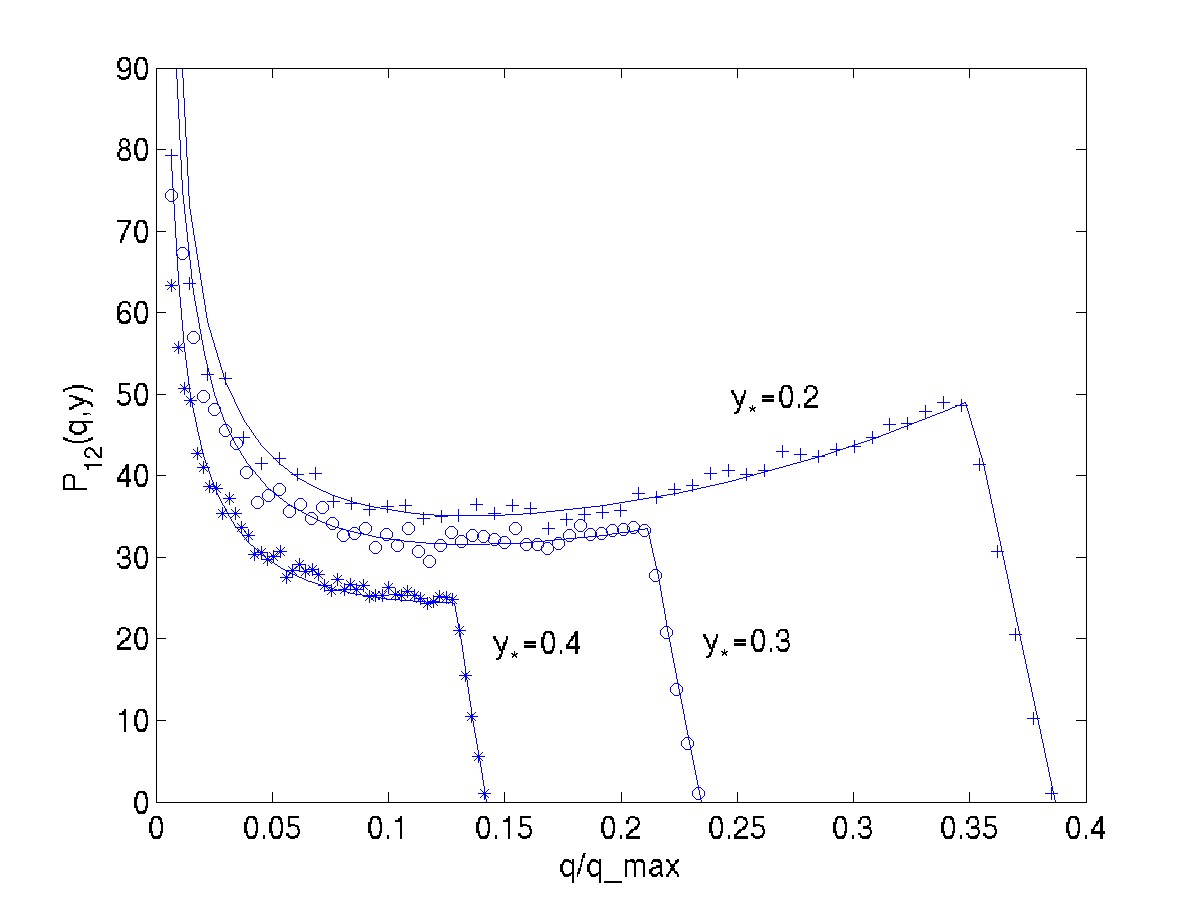}
\includegraphics[width=8cm,height=7cm]{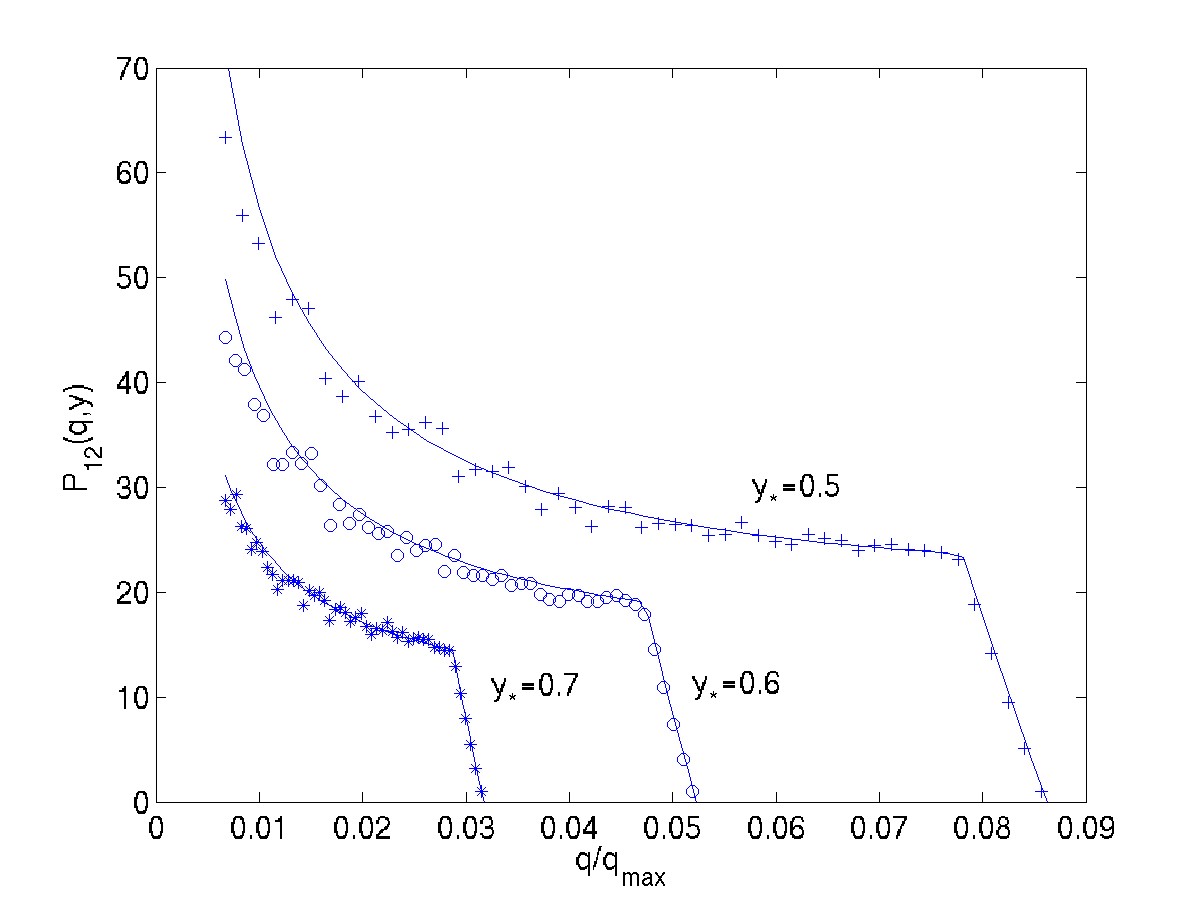}
\caption{\label{fig1} Comparison between $P_{12}(q,y)$ from Monte Carlo with the smooth portion of the analytic result
in \eqref{strip}. In this illustration $\alpha_*=5$. Crosses, open circles and stars are the Monte-Carlo \pdf{}s 
while the smooth solid curves are the analytical expression. 
Top panel: The bimodal \pdf{}s for $y_*<2/\alpha_*$. Middle panel: The \pdf{}s for larger $y_*$ which show the transition from a bimodal to a unimodal
distribution. 
Bottom panel: The unimodal \pdf{}s far from the source, i.e.\ for $y_*>2/\alpha_*$.}
\end{figure}

The expression in \eqref{B8.1} and \eqref{B8.2} provides the \pdf{} at a specified $y$. From an observational and  numerical view, recording the \pdf{} at a particular 
$y$ can be difficult. It is more likely that one obtains an estimate of the \pdf{} over a strip, i.e.\ for $y \in [y_1,y_2]$.
Integrating (\ref{B8.2}) over such a strip, and denoting this by $P_{12}(q)$, we have
\begin{align}
P_{12}(q) &= \int_{y_1}^{y_2} \!\!\!\! P(q,y) \, \dd y \, , \nonumber \\ 
&= \left[ \frac{\delta(q-\qmin )}{2L^2} + \frac{\alpha}{2Lq \log^2(q/\qmax)} \right]  
 s_{12}(q) \, ,
 \label{strip} 
\end{align}
where the strip function is
\begin{equation}
s_{12}(q) =\begin{cases} y_2^2-y_1^2\, , & \text{if $\qmin <q< \qs(y_2)$;} \\
y^2_s(q)-y_1^2\, ,  & \text{if $\qs(y_2)  <q< \qs(y_1)$.}
\end{cases}
\end{equation}
The slightly complicated form of  $s_{12}(q)$ is because when $q_s(y_2) < q < q_s(y_1)$,
the  upper limit of the integral is $y_s(q) < y_2$. We compare $P_{12}$ above with the results of a Monte Carlo simulation in Figure \ref{fig1}.

The effect of the strip function $s_{12}$ is to clip the   secondary peak  at $q=\qs$, and this clipping becomes more extreme if $y_2-y_1$ is made larger. Thus, aggregating measurements of $q$  from different spatial locations will obscure the bimodal structure of $P(q,y)$.

\subsection{The global \pdf{} of $q$}

In the extreme case $y_1 \rightarrow 0$ and $y_2 \rightarrow L$, we obtain the global \pdf{} of the specific humidity, denoted by  $g(q)$, in \eqref{global1}. Using \eqref{global2}, the global \pdf{} in this example is
\begin{equation}
g(q) = \frac{1}{2} \delta(q - \qmin) + \frac{1}{2 \alpha L q }\, ,    
 \label{avgpdf} 
\end{equation}
where $q\in [\qmin,\qmax]$.
The smooth part of the global \pdf{} is  $g(q) \sim q^{-1}$.

\subsection{Monte Carlo Simulations}

To test the analytic solution above, we proceed to a Monte Carlo simulation of our system. Typically our simulations use $N=10^5$ parcels which are released in the domain 
$y \in [0,L]$. We have used $\alpha=1, L=5$ which gives $\alpha_*=5$. This implies a change of approximately two orders of magnitude 
in the saturation 
specific humidity across the domain. The $y$-coordinate of the  
parcels are initially uniformly distributed and are tagged with the minimum saturated specific humidity, i.e. on parcel $i$, $q_i (0) = \qmin$. The 
parcels perform an uncorrelated random walk in $y$, and parcel $i$  reflected back into the 
domain if $y_i > L$ or if $y_i < 0$. The forcing $\S$ is 
implemented by setting $q_i = \qmax$ when a parcel is reflected at $y=0$. Rapid condensation is implemented by enforcing $q_i=\min[q_i,q_s(y_i)]$ after each random displacement.

 To approximate brownian motion, the root mean square step length of the random walk, $\ell$, should be as small as computationally feasible --- we typically used  $\ell = 2 \times 10^{-3} L $. We specify the brownian diffusivity as $\kappa_b=0.5$, and  then the time between steps, $\tau$, is computed via
\begin{equation}
\kappa_b = \frac{\ell^2}{2 \tau}.
\end{equation}
The simulation runs till $P(q,y)$ attains a stationary form and then data is collected in several strips for comparison with \eqref{strip}.

\begin{figure}
\centering
\includegraphics[width=8cm,height=7cm]{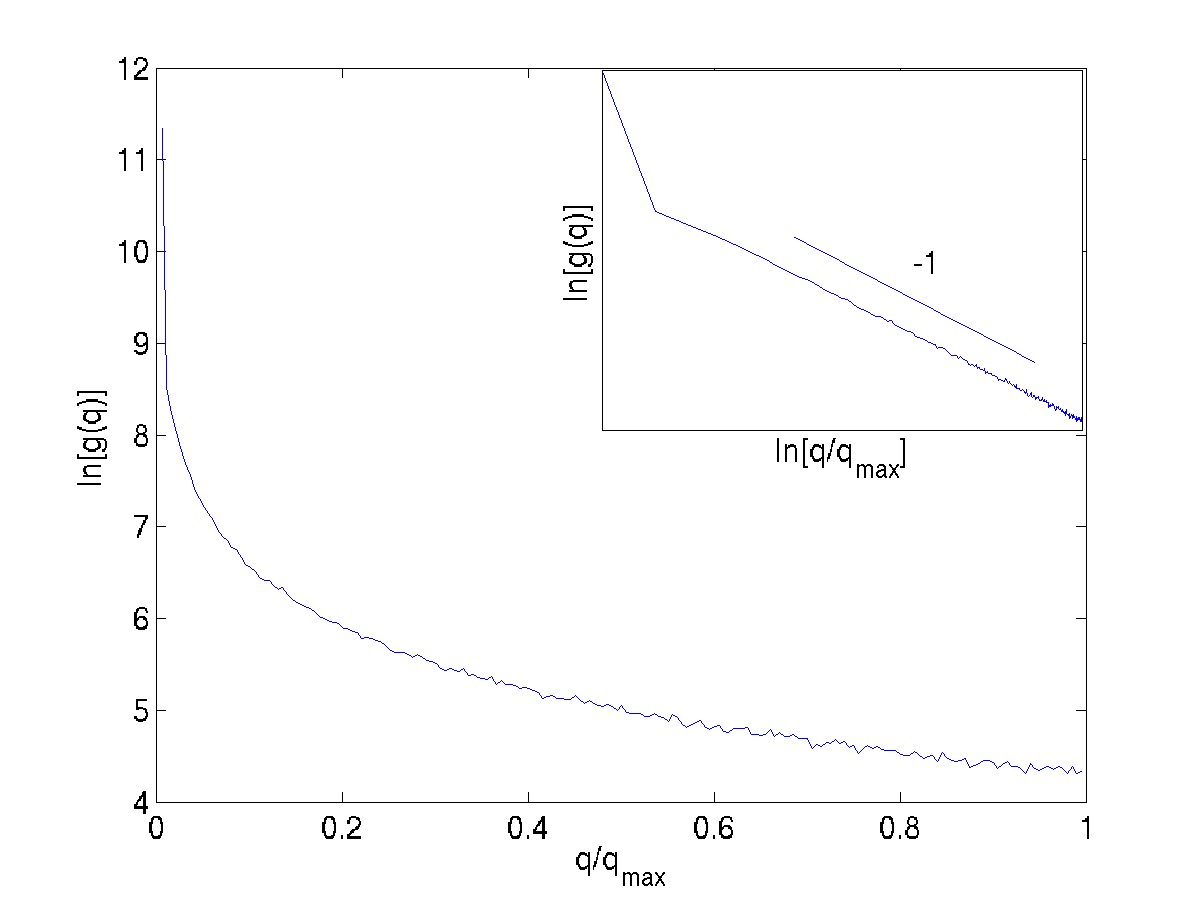}
\includegraphics[width=8cm,height=7cm]{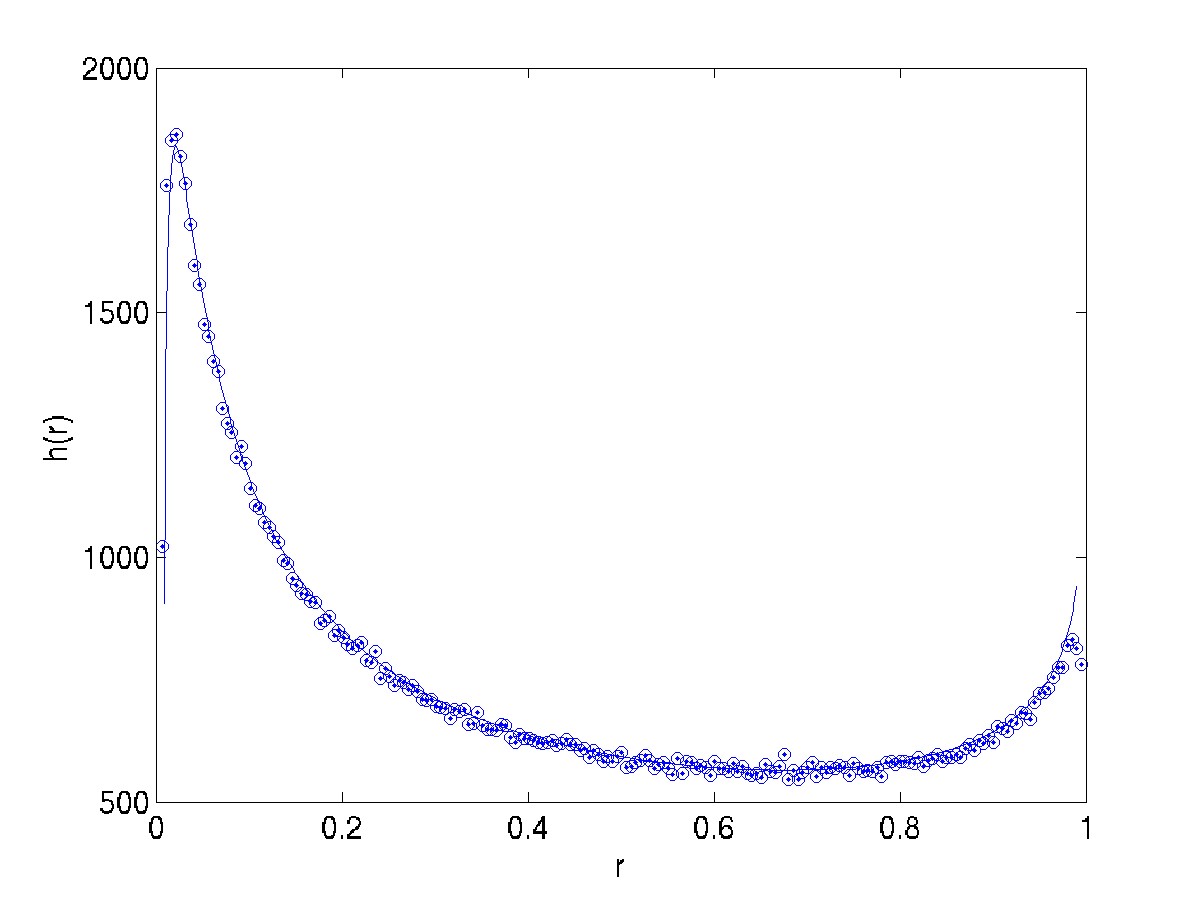}
\caption{\label{fig_zz} Top panel: The global \pdf{} of the specific humidity $g(q)$ from Monte Carlo simulations. The inset shows a $\log$-$\log$ plot
which reveals a spike at $\qmin$ followed by an algebraic $q^{-1}$ tail as predicted by \eqref{avgpdf}. Bottom panel: The analytical curve
$h(r)$ from \eqref{Bb} (solid line) and the global RH \pdf{} from the numerical simulation (dotted circles).}
\end{figure}

The results of the simulations are shown in Figure \ref{fig1}. The top panel of this figure shows a comparison between the smooth 
portion of the theoretical 
expression in \eqref{strip} and the numerically estimated \pdf{} for $y_*<2/\alpha_*$. 
There is  good agreement between the Monte-Carlo simulation and the Fokker-Planck solution. Some minor discrepancies are evident as $q \rightarrow \qmin$; 
this is anticipated as we have suppressed the numerical spike at
$q=\qmin$ (indeed, the theoretical \pdf{} is given by a $\delta$-function at this location). The middle panel of Figure \ref{fig1} shows the 
\pdf{} at larger $y$, specifically around the region where the \pdf{} transitions from a bimodal to a unimodal form.
The bottom panel of Figure \ref{fig1} shows the \pdf{}'s far from the source, where the \pdf{}s have  a single peak at $q=\qmin$, 
followed by decay and termination at $q=q_s(y)$. 
Finally, the first panel of Figure \ref{fig_zz} shows
the global \pdf{} $g(q)$ in \eqref{avgpdf}. As per the $\log$-$\log$ inset,  the global \pdf{} closely follows the result
in \eqref{avgpdf}, with a spike at $q=\qmin$ followed by an algebraic  tail  $g(q) \sim q^{-1}$.

\subsection{Transport and condensation}

To  understand the   water vapour distribution, it is important to quantify the 
transport of moisture.
In the context of the AC model, it was shown by \cite{GS-2006} that brownian motion results in the mean moisture flux
\begin{equation}
\langle \F \rangle  = -\kappa_b\, \frac{\dd \langle q \rangle}{\dd y}\, ,
\label{m1}
\end{equation}
where $ \langle q \rangle$ is the mean specific humidity
and $\kappa_b$ is the brownian  diffusivity. 
Then, as in \eqref{condy11}, in a statistical steady state the convergence of $\langle \F \rangle$  balances the mean condensation $\la \C\ra$ at every $y$. 

\begin{figure}
\centering
\includegraphics[width=8cm,height=7cm]{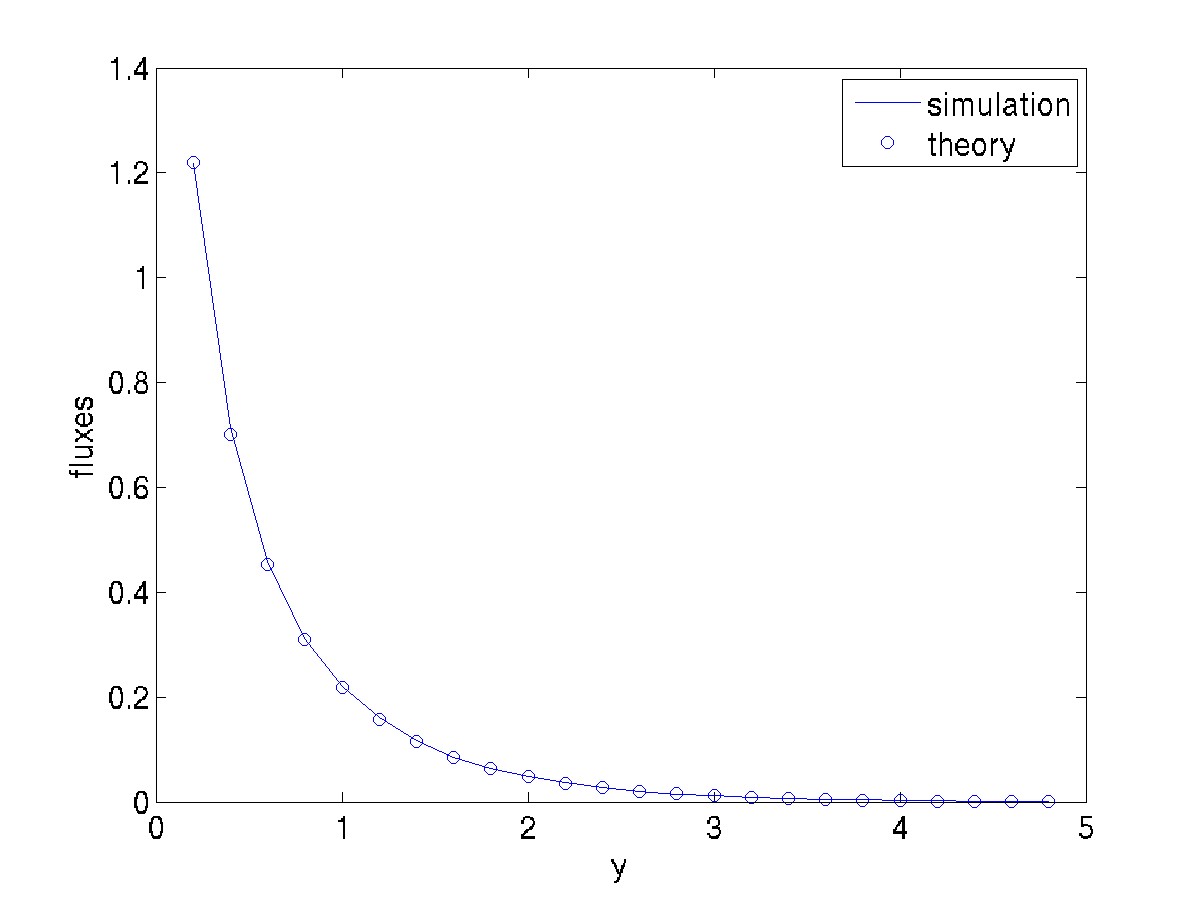}
\caption{\label{fig1a} A comparison of the flux $\la \F\ra$ measured in the  Monte Carlo simulation with the theoretical result using \eqref{m1} and \eqref{m2}.} 
\end{figure}

Using $f(q)=q$ in  \eqref{avg2}, and $\qs(y)$ in \eqref{exp},  the average gradient of specific humidity is obtained as 
\begin{equation}
\frac{\dd \langle q \rangle}{\dd y} =  \alpha \qmax \left[\mathrm{E}(\alpha L) - \mathrm{E}(\alpha y) \right]\, ,
\label{m2}
\end{equation}
where $\mathrm{E}(x) \defn \int_x^{\infty} (\ee^{-t}/t) \,  \dd t$ is the exponential integral. 
The moisture flux $\langle \F \rangle$ can also be diagnosed from the simulation and thus \eqref{m1} is tested with  agreement shown in  Figure \ref{fig1a}.

\subsection{The Relative Humidity (RH)}

It is the immense subsaturation of the troposphere that makes the study of water vapour an interesting and 
challenging problem \citep{SB, PBR}. 
This subsaturation is most starkly revealed by examining the RH, which is
\begin{equation}
r \defn \frac{q}{\qs(y)}\, .
\end{equation}
Converting the \pdf{} in \eqref{B8.2} to a \pdf{} for $r$, denoted by $Q(r,y)$, gives
(for $y > 0$)
\begin{equation}
Q(r,y) = \frac{y}{L} \left[ \frac{\delta( r-\rmin) }{L}+ \frac{\alpha}{r \log^2(\epsilon r/\rmin)} \right]\, ,
\label{B10}
\end{equation}
where  $r \in [\rmin(y),1]$,  $\epsilon$ is defined \eqref{epsilondef},  and
\begin{equation}
\rmin(y)  = \epsilon \ee^{\alpha y} \, .
\end{equation}

\begin{figure}
\centering
\includegraphics[width=8cm,height=7cm]{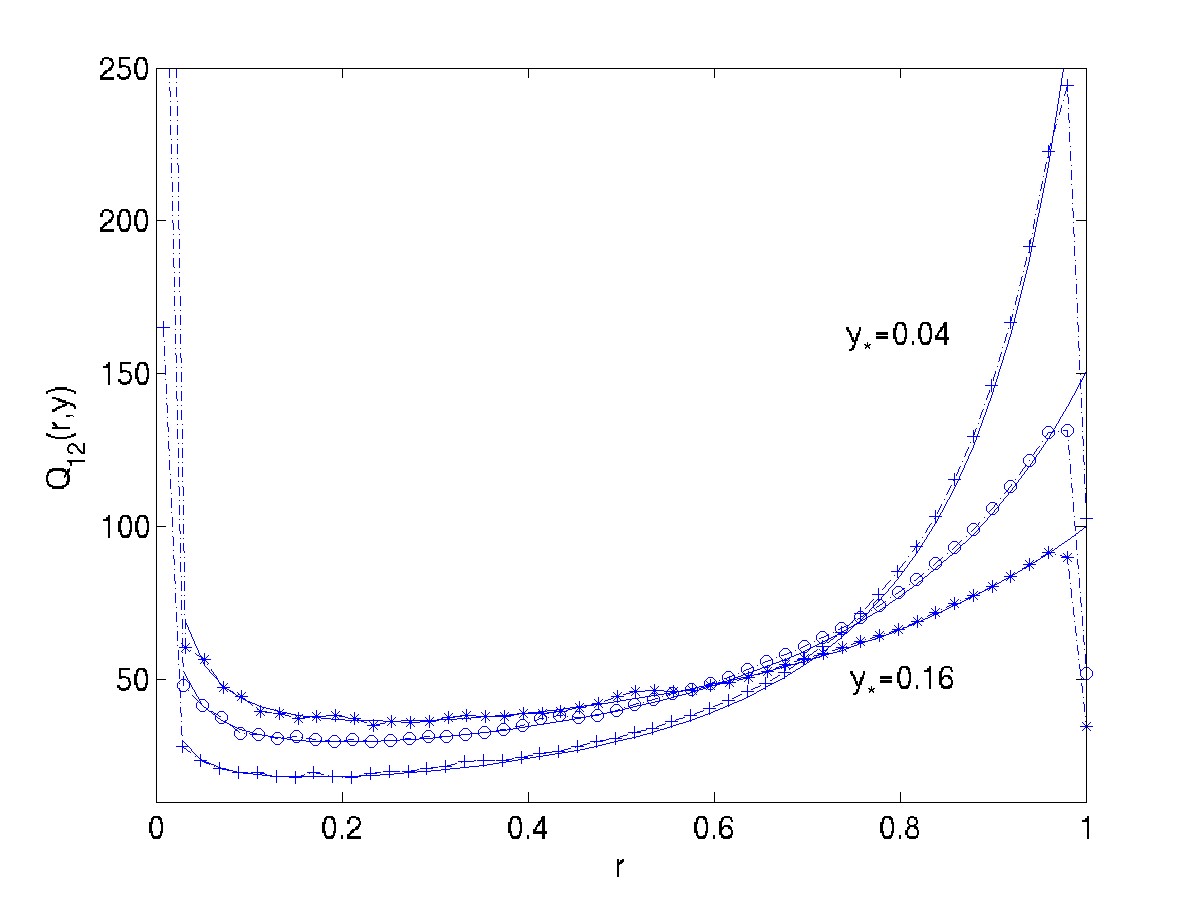}
\includegraphics[width=8cm,height=7cm]{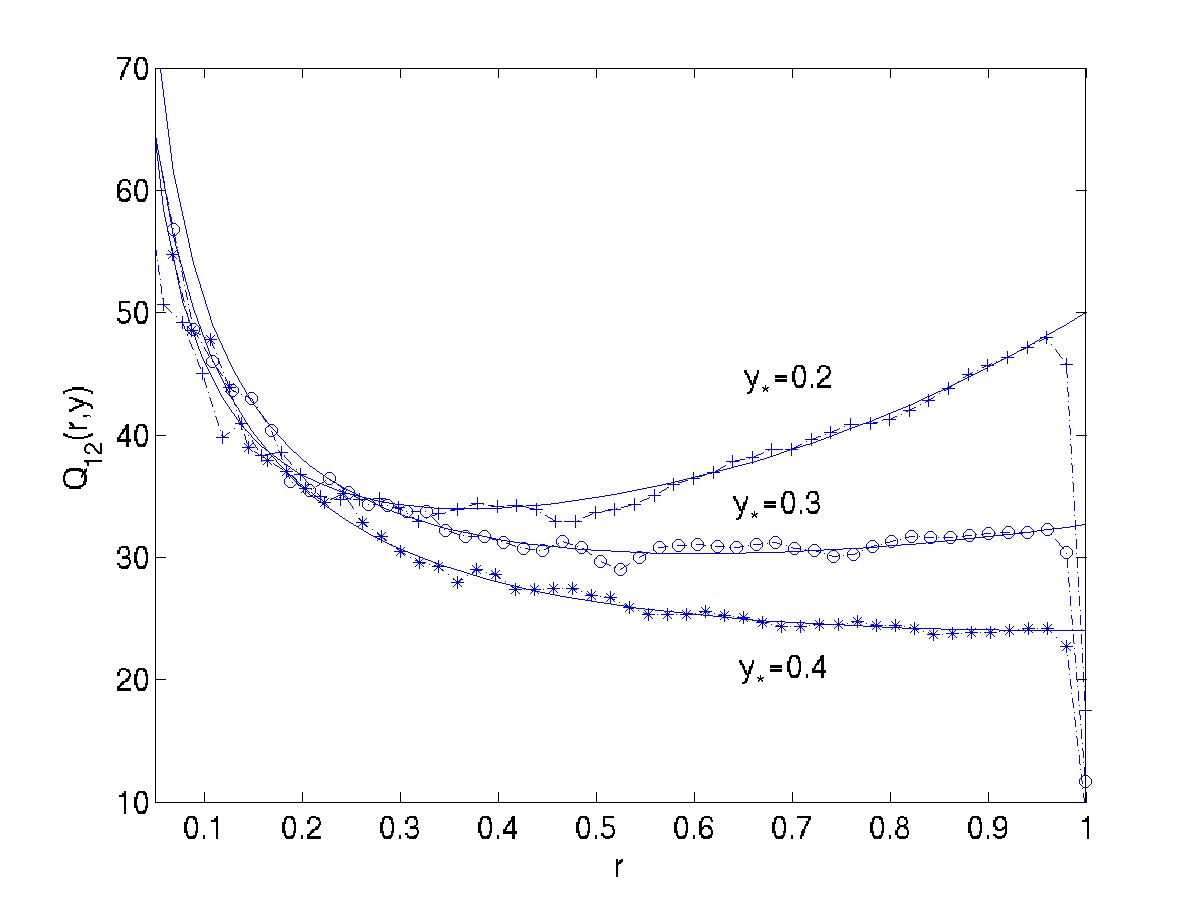}
\includegraphics[width=8cm,height=7cm]{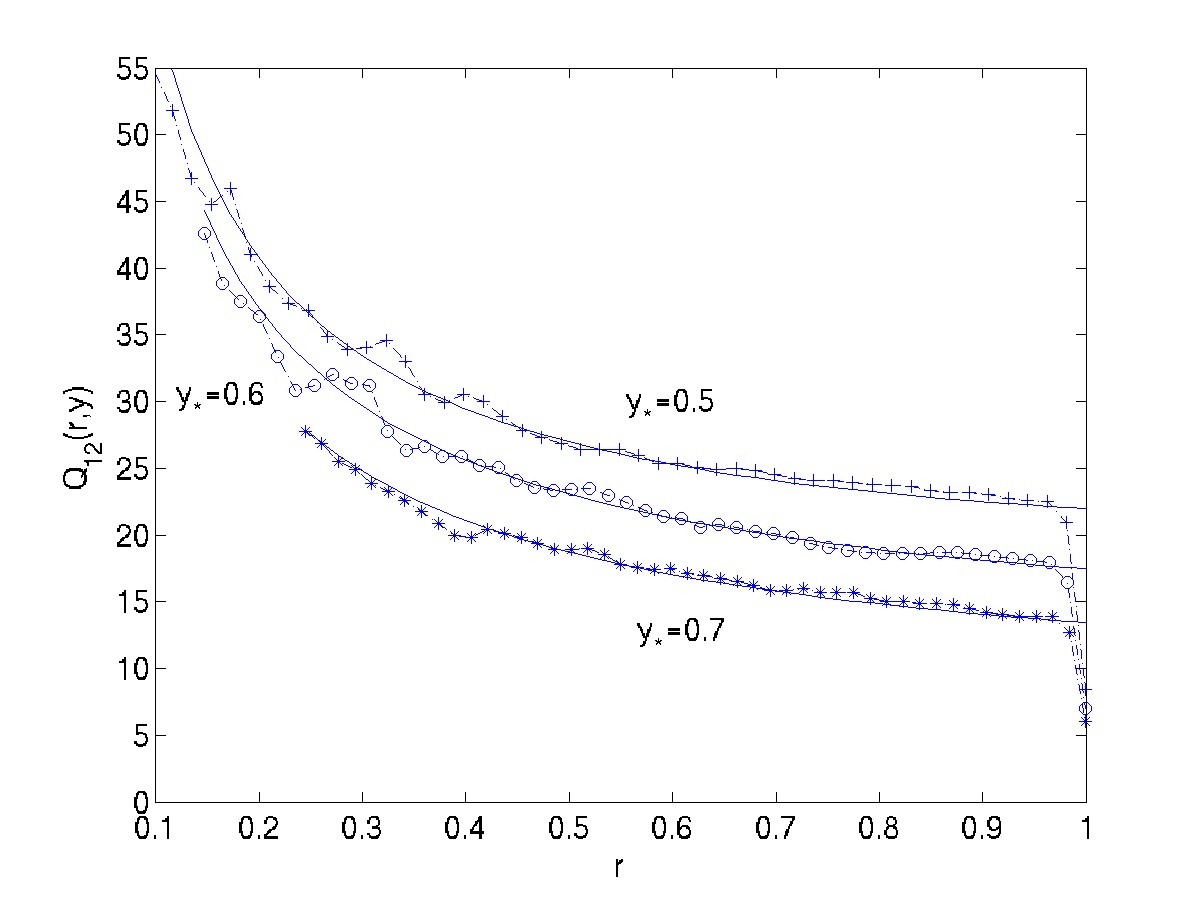}
\caption{\label{fig2} 
Comparison between $Q_{12}(r,y)$ from Monte Carlo with the smooth portion of the analytic result
in \eqref{B10}. In this illustration $\alpha_*=5$  and  $y_2 - y_1 = 0.02 L$. Crosses, open circles and stars are the Monte-Carlo \pdf{}s
while the smooth solid curves are the analytical expression.
Top panel: The bimodal \pdf{}s for $y_*<2/\alpha_*$ (the curve in the middle corresponds to $y_*=0.08$). 
Middle panel: The \pdf{}s for larger $y_*$ which show the transition from a bimodal to a unimodal
distribution.
Bottom panel: The unimodal \pdf{}s far from the source, i.e.\ for $y_*>2/\alpha_*$.}
\end{figure}

A comparison between (\ref{B10}) and the Monte Carlo simulation is shown in Figure \ref{fig2}.
The main mismatch we see is for $r \rightarrow 1$, where the numerical \pdf{}'s
taper off while expression (\ref{B10}) continues to rise till $r=1$. 
This discrepancy is due to the collection of numerical data in a strip (as before), while
\eqref{B10} provides an expression for the \pdf{} at a particular location.
Note that
the numerical \pdf{}'s show a spike at
$\rmin(y)$ --- shown only in the first panel --- which agrees with the shifting $\delta$-function in (\ref{B10}). 
Bimodality of the \pdf{} at small $y$ is evident, while farther from the source there is  a peak
at $\rmin(y)$ (suppressed in the plot), after which the \pdf{} gradually levels off as $r \rightarrow 1$. The transition from a bimodal to 
a unimodal \pdf{}, as well as the \pdf{}s far from the source, are shown in the
second and third panels of Figure \ref{fig2} respectively.

\subsection{The global RH \pdf{}}

The 
global \pdf{}, $h(r)$, of RH is 
\begin{equation}
h(r) \defn \int_0^{ \log \left( r/\epsilon \right)/\alpha} \!\!\!\!\!\!\!\!\!\!\!\! Q(r,y) \dd y .
\label{Ba}
\end{equation}
Evaluating \eqref{Ba} using $Q(r,y)$ in  \eqref{B10}
gives\begin{equation}
h(r) = \frac{1}{\alpha L r}  \log \left[ \frac{\alpha L}{\log(1/r)} \right] .
\label{Bb}
\end{equation}
As a check one can verify that on integration over the range of RH, that is $\epsilon \leq r \leq 1$, the global \pdf{}  in \eqref{Bb} normalizes to unity.
The lower panel of Figure \ref{fig_zz} shows a comparison between the Monte Carlo global RH \pdf{} and \eqref{Bb}. There is a good match between the theoretical prediction \eqref{Ba} and the simulation.
The global \pdf{} exhibits a mix of the distinct features of the \pdf{}s at different $y$ e.g., $h(r)$ is bimodal with a primary peak at small RH,
followed by a decay at intermediate RH, and then a
secondary ``moist peak'' as $r \rightarrow 1$.

\section{Discussion \label{discussion}}

\subsection{Radiative effects of water vapour}

\begin{figure}
\centering
\includegraphics[width=8cm,height=7cm]{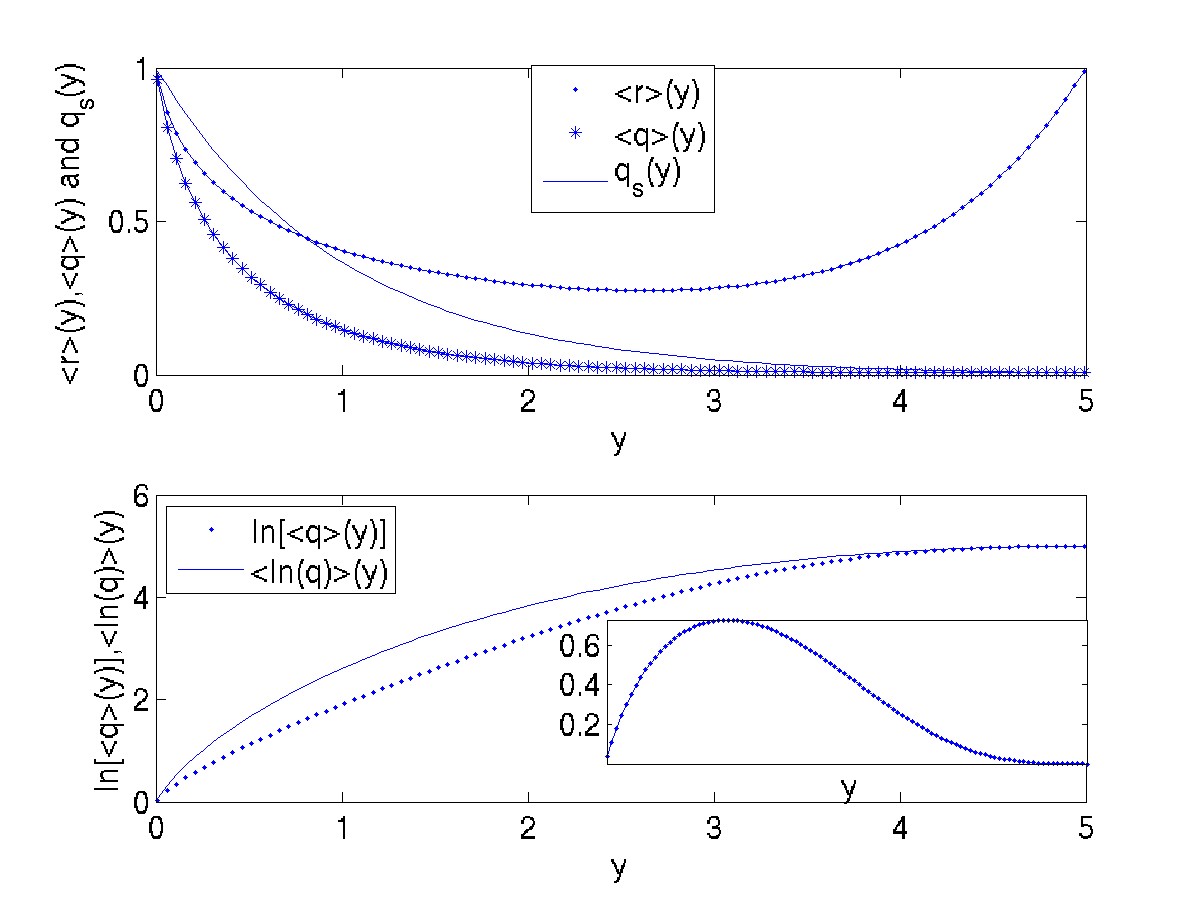}
\caption{\label{fig3} Upper panel: The 
average profiles of the specific humidity and RH  computed from the \pdf{}s in (\ref{B8.2}) and (\ref{B10}) respectively. Lower panel: $\log \langle q \rangle (y)$
and $\langle \log(q) \rangle (y)$ as computed using \eqref{avg1}. The inset shows the difference $\langle \log(q) \rangle (y) - \log \langle q \rangle (y)$ and demonstrates the 
increase in the OLR due to fluctuations in $q$ at every $y$.}
\end{figure}

In the introduction, we referred to the importance of the vapour \pdf{} in understanding the effect of humidity on the OLR. 
In fact, the  effect of humidity on OLR was quantified by
 \cite{ZSM} in 
terms of the \pdf{} of the RH field. Their numerical experiments, employing a full-fledged radiation code, showed that a domain with a bimodal 
RH \pdf{} allowed for significantly larger cooling than one with 
a unimodal distribution (holding the total and mean the same in the two cases). In the present context, we see that the \pdf{} of the RH over the 
domain, i.e.\ the lower panel of 
Figure \ref{fig_zz}, is bimodal. Therefore, along with regions of moderate RH, the domain consists of subsets with very small RH 
(the first peak in the aforementioned plot)
and those that are nearly saturated (the second peak). A clearer picture is seen in the upper panel of Figure \ref{fig3} which shows the average
specific humidity and RH over the domain, i.e.\ $\langle q \rangle (y)$ and 
$\langle r \rangle(y)$ 
respectively. 
While the specific humidity decays monotonically from the source, 
the RH has a pronounced mid-domain minimum representing a pool of ``relatively'' dry
air which is flanked on either side by nearly saturated regions. 

As described in PBR, the presence of water vapour affects the OLR in an approximately logarithmic fashion, i.e.\ OLR $\propto -\log(q)$. Therefore, the importance of 
knowing the \pdf{} of $q$ at every $y$ is illustrated by comparing 
$\log( \langle q \rangle)$ 
with  $\langle \log(q) \rangle $ --- it is the latter quantity that accounts for fluctuations in the specific humidity at every location and requires a knowledge of $P(q,y)$. 
As expected (see for example the discussion in PBR), and as is shown 
in the lower panel of Figure \ref{fig3}\footnote{Both curves start at zero as we have chosen $\qmax=1$, in general the value at $y=0$ is $\log(\qmax)$.}(especially the inset, which shows the difference between the two estimates)
the  fluctuations in the specific humidity field increase the OLR as 
compared to the estimate based only on the mean profile $\langle q \rangle (y)$.

\subsection{Bimodality of the RH \pdf}

It is interesting that the global \pdf{} of $q$ in the upper panel of Figure \ref{fig_zz} is unimodal, while the global \pdf{} of RH in the lower panel of Figure \ref{fig_zz} is bimodal. Thus, according to our model, RH bimodality  is so robust that it survives the mixing of  \pdf{}s from diverse locations into the global \pdf{} $h(r)$ in \eqref{Bb}.

As far as we are aware, bimodal RH \pdf{}s
that have been documented in literature are mostly derived from satellite data over the deep tropics \citep{BZ-1997, ZSM, MF-2006, LKJ-2007}. Exceptions are the three-dimensional and isentropic model based AC experiments by
PBR and \cite{YP-1994} respectively: in Figure 16 in PBR and
in Figure 6 of \cite{YP-1994} the midlatitude RH \pdf{} is clearly bimodal.

On the other hand,  the stochastic drying models developed by \cite{SKR} and \cite{Ryoo} predict only unimodal RH \pdf's. 
Indeed, \cite{SKR} concluded that a broad and unimodal RH density might be regarded as a general outcome of the AC model. 
This conclusion is inconsistent with our implementation of the AC model, which results in bimodal RH \pdf{}s. We discuss this point further in section \ref{robust}.

Because our main focus is humidity fields on middleworld isentropic surfaces,  we proceed with a  data comparison by 
analyzing the RH on the 330K isentropic surface as presented in the ERA interim product. 
The data
was obtained from the ECMWF web portal (\url{http://data-portal.ecmwf.int/data/d/interim_daily/}). We analyse daily data for the year 2008 given at a horizontal resolution of
$1.5^\circ \times 1.5^\circ$ and at 24 vertical levels from 1000 mbar to 175 mbar. As the RH data is only available on isobaric levels, we interpolate
to generate RH fields on the 330K isentrope. Further, as we are interested in the \pdf{}s from the midlatitudinal portion of the isentropic surface, we
collect data between $15^\circ$ and $60^\circ$ in both the hemispheres. 

\begin{figure}
\centering
\includegraphics[width=8cm,height=7cm]{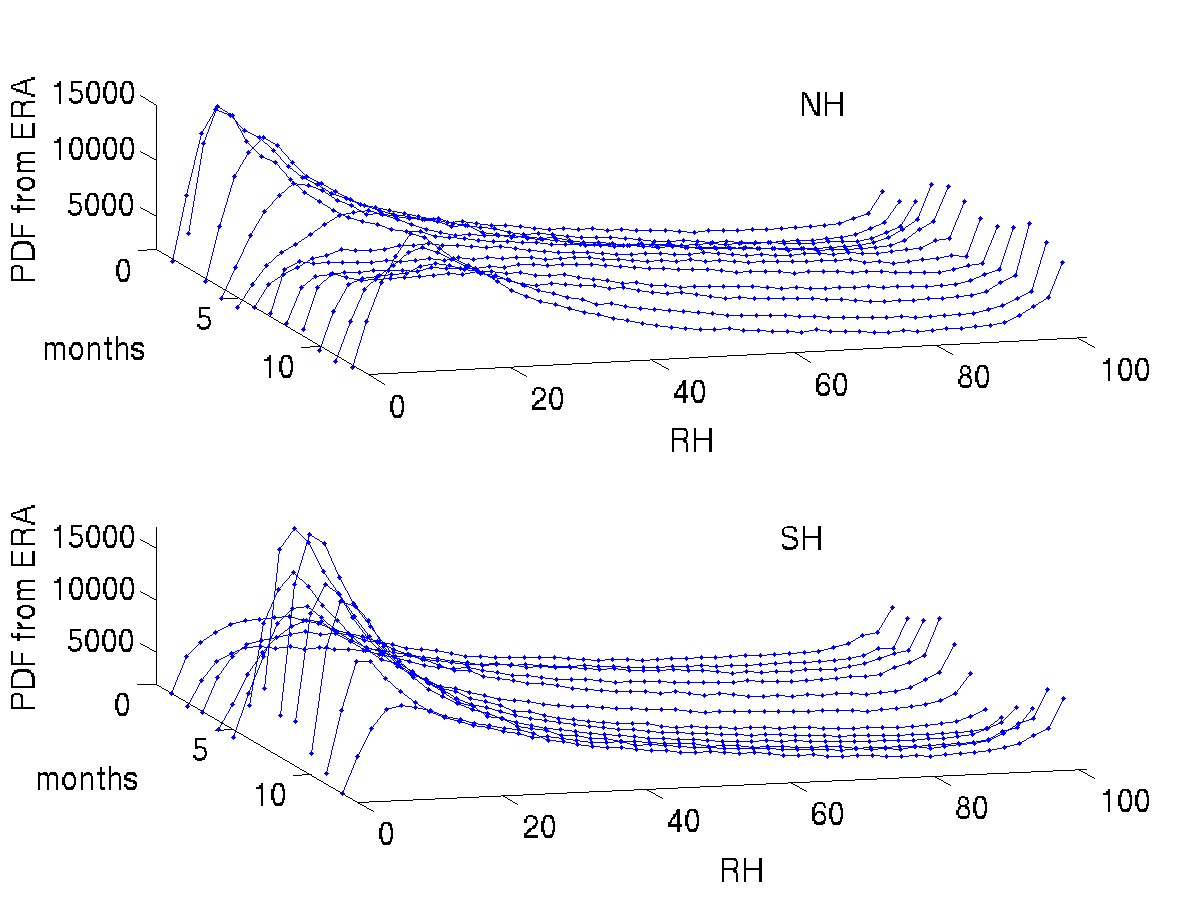}
\includegraphics[width=8cm,height=7cm]{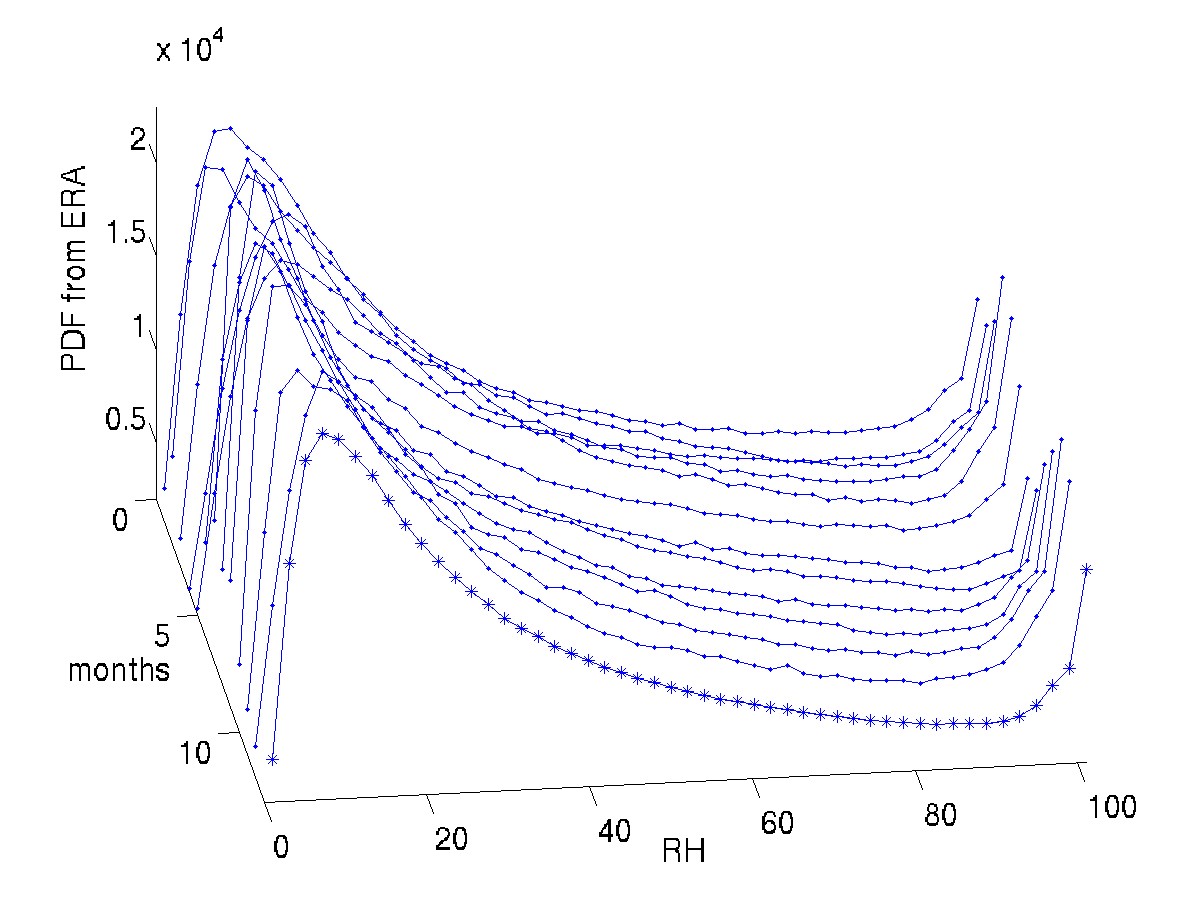}
\caption{\label{fig3a} The midlatitudinal RH \pdf{} on the 330K isentrope as derived from the ERA interim data product for the individual months of 2008. 
Upper panel: \pdf{}s from the 
two hemispheres plotted separately (ordering: January --- back to December --- front). Lower panel: 
The lines with dots are the \pdf{}s with both hemispheres taken together, and the foremost
bold line with star symbols is the mean of these monthly curves.}
\end{figure}

A reviewer noted that the 330K 
isentrope intersects the tropopause by $60^\circ$. To alleviate concerns that these bimodal \pdf{}s result from sampling distinct air masses separated 
by a mixing barrier, 
we have also constructed \pdf{}s by considering
data only from the troposphere, for example, from $15^\circ$ and $35^\circ$ in both hemispheres, and verified that these \pdf{}s display the same 
qualitative form as those  in Figure \ref{fig3a}.

As seen in Figure \ref{fig3a} --- which shows the \pdf{}s from the different months of 2008 (\pdf{}s from the individual hemispheres are plotted in 
the upper panel, and a composite from both the hemispheres is shown in the
lower panel) ---
there are considerable differences in the RH \pdf{}s between the two hemispheres. For example, 
the dry peak in the \pdf{}s is more prominent in each hemisphere's winter season.
As these differences are not of primary interest to us we do not dwell on them here. 
There are a few values in the midlatitudes where 
the ERA dataset shows supersaturation,
these have been set to $100 \%$ in the present calculation. Naturally, this raises the secondary peak but does not not alter the form of the \pdf{}.
Quite clearly, the \pdf{} does vary from month to month but its main features are robust throughout the year. 
More importantly, we observe that the 
ERA interim data \pdf{}, in all months, is bimodal with distinct dry and moist peaks, which conforms with the results of the 
AC model in section \ref{example}.

\subsection{Special features of the AC model  in this paper\label{robust}}

In reconsidering the different ingredients in our idealized implementation of the AC model, we emphasize that there are at least four important  assumptions affecting our conclusions:
\begin{itemize}
\item Isentropic transport is modeled as  brownian motion;
\item Parcels are re-moistened only at  $y=0$;
\item The saturation humidity is    $\qs(y)=\qmax \ee^{-\alpha y}$.
\item The model neglects diffusive processes.
\end{itemize}

The first assumption, which is equivalent to saying that the lagrangian velocity correlation time is zero,  is crucial for the derivation of the Fokker-Planck equation, 
and thus for all of the results in this paper. Moreover, following \cite{GS-2006}, we emphasize that the important condensation-diffusion balance  in \eqref{condy11} 
also relies crucially on the assumption of brownian motion. 
With nonzero velocity correlation time it might still possible to argue that mean condensation $\la \C\ra$ is balanced by diffusive convergence, that is
\begin{equation}
\la \C \ra = \kappa_q \frac{\dd^2 \la q\ra}{\dd y^2}\, .
\end{equation}
But the humidity diffusivity $\kappa_q$ will be systematically less than the brownian parcel diffusivity $\kappa_b$. Understanding the relation between $\kappa_q$ and $\kappa_b$ is beyond our scope here.

The second assumption of a localized vapour source probably explains the difference between our conclusions and those of \cite{SKR} regarding bimodality of the RH \pdf{}s. 
This point could be examined within the Fokker-Planck framework by modeling the source $\S$ in \eqref{1a} as a re-setting, $q \to \qs(y)$, occurring at a 
rate $\tmoist(y)$ throughout the domain. We do not pursue this generalization here.

The third assumption, that the saturation humidity varies exponentially with $y$, is a consequence of the Clausius-Clapeyron relation and the assumption that the temperature 
gradient on a isentrope is uniform. We have examined alternative models of $\qs(y)$, e.g., taking $\qs(y)$ as a linear or a parabolic function of $y$ 
while ensuring $\qs(y)$ is still monotonically decreasing $\qmax$ at $y=0$.
The main qualitative features of these solutions shown in 
Figures \ref{fig1}, the lower panel of Figure \ref{fig_zz} and 
in Figure \ref{fig2} are unchanged. 
The only \pdf{} which is sensitive to $\qs(y)$ is the global specific humidity \pdf{}, $g(q)$ shown in  the upper panel of Figure \ref{fig_zz}. Thus, as compared to
the $q^{-1}$ algebraic decay in the exponential $\qs$, a linear (parabolic) saturation $\qs(y)$ yields a uniform (growing) \pdf{} as $q \rightarrow \qmax$. These differences can be traced to the form of the function $\ys(q)$ in the analytic expression for $g(q)$ provided by \eqref{global2}.

The fourth assumption above is also crucial to the AC model.  \cite{HA97} estimate that in the upper troposphere and lower stratosphere  the ``mix-down" time for synoptic-scale fluid parcels to be sheared and strained  to molecular-diffusive scales is a few weeks. 
This time scale is comparable to the mean residence time of water in the free troposphere \citep{Quante}. Thus molecular diffusion between fluid parcels may have an 
impact on the water vapour field, and in particular on its \pdf{}. 
This  important open question cannot be addressed till the advection-condensation model is extended to the advection-diffusion-condensation model.

\section{Conclusions \label{conclusion}}

We have studied the AC model forced by resetting the specific humidity at the southern boundary of the model isentrope. 
If the flow can be approximated by brownian motion (i.e., a very rapidly de-correlating velocity), then the \pdf{} of the vapor field is governed by the Fokker-Planck equation, which can be solved in the limit of rapid condensation. 
The solution is illustrated for a saturation profile which decays exponentially along the isentrope in the meridional direction.
Analysis of this solution, and supporting  Monte Carlo simulations, document a non-trivial statistical steady state in which the southern moisture source 
is balanced by condensation throughout the interior of the domain. ``Non-trivial'' refers to
the fact that the steady state is far from the limiting states of complete saturation or complete dryness, and instead
is characterized by a  spatially inhomogenous \pdf{}. This non-uniformity manifests itself in the co-existence of very 
dry and nearly saturated regions within the domain. 
We find this to be encouraging as a strongly inhomogeneous steady state is reminiscent of the  atmosphere. In fact, the bimodality of the 
global RH  \pdf{} predicted by the model is supported by  
midlatitudinal middleworld isentropic RH \pdf{}s from the 
ERA interim product. 

An important property of our solution is its spatial structure: near the resetting source (i.e.\ for small $y$) 
the \pdf{} is bimodal with a peak at $q_{\textrm{min}}$, a rapid decay
followed by a rise to a second peak as $q \rightarrow q_s(y) \approx \qmax$. Futher  from the source (i.e.\ for large $y$) the \pdf{} 
is unimodal and terminates at $q = \qs(y) \ll \qmax$. To facilitate a comparison with 
numerical experiments and observations we also derived an expression for the \pdf{} as estimated over a strip of non-zero width. Extending the 
strip to cover the domain, we obtained the domain averaged \pdf{} which in the case of an exponential saturation profile has a $q^{-1}$-form. These features
are in very good agreement with numerical simulations employing a large number of diffusively moving parcels. We also demonstrated the 
utility of the \pdf{} in estimating 
mean field observables such as the moisture flux and condensation rate. 

From the \pdf{} of the specific humidity we deduced the  \pdf{} of the RH. Once again, the \pdf{}s near the 
source are bimodal with a peak at the minimum RH for a given $y$, i.e.\ at $\rmin(y)$, followed by a decay at intermediate 
RH values and a rapid rise as $r \rightarrow 1$. The bimo      a unimodal distribution at large distances from the 
source, and here the \pdf{} is controlled by the $\delta$-function at $\rmin(y)$ which levels off to a constant as $r \rightarrow 1$.
The domain averaged RH \pdf{} displays a mix of these distinct features, 
in particular --- even though the global specific humidity \pdf{} is unimodal --- the global RH \pdf{} is bimodal with a primary peak at small RH,
followed by a decay at intermediate RH, and then a
secondary peak as $r \rightarrow 1$. 

Though this work deals with an idealized model, we believe the  analysis afforded in this setting provides a useful understanding of
advection-condensation and the equilibria supported by the interaction of these processes with a steadily maintained source of moisture. 
Indeed, with a proper identification of the saturation profile and the resetting protocol, our methodology can be applied wherever the AC model finds use. 
Also, we argue
that the algorithm outlined here is of practical interest in that, given a saturation profile from a general circulation model, one can then predict the 
form of the \pdf{} of the vapour and relative humidity field. In turn, this can be compared with the \pdf{}s generated directly from the model itself, 
thus providing a method to validate the 
fidelity of moisture evolution in a general circulation model. 

\ack WRY is supported by the NSF  under Grant No.~OCE07-26320. We thank Paul O'Gorman for a useful discussion of the AC model.

\appendix

\section{Normalization as a differential boundary condition}

Here we show that the normalization constraint obtained in the limit of rapid condensation can be interpreted as differential boundary condition. 
Specifically, we consider
this interpretation for the
resetting problem and pick up the story at \eqref{5} where 
\begin{equation}
P(q,y) = \frac{\Phi(q)}{L} + \left[ \frac{\delta(q-\qmin)}{L^2} + F(q)\right] y\, .
\label{Ap1}
\end{equation}
Applying the normalization \eqref{prominent}, for $y \in (0,L)$ we obtain
\begin{equation}
\frac{y}{L^2} + y \int_{\qmin}^{\qs(y)} \!\!\!\!\! F(q) \, \dd q = \frac{1}{L}.
\label{Ap2}
\end{equation}
Differentiating with respect to  $y$, Leibnitz's rule gives
\begin{equation}
\frac{1}{L^2} + \int_{\qmin}^{\qs(y)} \!\!\!\!\! F(q) \dd q + y  F[q_s(y)]\,  \frac{\dd q_s}{\dd y}  = 0\, .
\label{Ap3}
\end{equation}
A second differentiation yields,
\begin{equation}
2 F[q_s(y)] \frac{\dd q_s}{\dd y} + y \frac{\dd F}{\dd y} \left( \frac{\dd q_s}{\dd y} \right)^2 + y F[q_s(y)]
\frac{\dd^2 q_s}{\dd y^2} = 0\, .
\label{Ap4}
\end{equation}
On specifying $q_s(y)$, we see that (\ref{Ap4}) is an ordinary differential equation for the function $F(q)$. Indeed, (\ref{Ap4}) is the
interpretation of the normalization constraint as a differential boundary condition. To see the utility of (\ref{Ap4}) in solving for $F(q)$,
consider $q_s(y)=\textrm{e}^{-\alpha y}$, this
yields
\begin{equation}
[2+\log(x)]F(x) + x \log(x) \frac{dF(x)}{dx} = 0.
\label{Ap5}
\end{equation}
and hence, $F(q)={\alpha}/{(L q \log^2(q))}$ which agrees with the earlier result (\ref{B8.2}).

\end{document}